\documentclass{aa}
\usepackage{graphics}

\newcommand{\am}[2]{$#1'\,\hspace{-1.7mm}.\hspace{.0mm}#2$} 
\newcommand{\HI}{\mbox{H\,{\sc i}}}

\begin{document}      

   \title{NGC~4569: recent evidence for a past ram pressure stripping event}

   \author{B.~Vollmer\inst{1,2}, C.~Balkowski\inst{3}, V.~Cayatte\inst{3}, W.~van\,Driel\inst{3}, \& W.~Huchtmeier\inst{2}}

   \offprints{B.~Vollmer, e-mail: bvollmer@astro.u-strasbg.fr}

   \institute{CDS, Observatoire astronomique de Strasbourg, UMR 7550, 11, rue de l'universit\'e, 
     67000 Strasbourg, France \and
     Max-Planck-Institut f\"ur Radioastronomie, Auf dem H\"ugel 69, 53121 Bonn Germany \and
     Observatoire de Paris, Section de Meudon, GEPI, CNRS UMR 8111 and Universit\'e Paris 7, 
     5 place Jules Janssen, 92195 Meudon Cedex, France
   } 
          
   \date{Received / Accepted}

   \authorrunning{Vollmer et al.}
   \titlerunning{NGC~4569}

\abstract{
Deep 21-cm \HI\ line observations of the Virgo cluster spiral galaxy NGC~4569 have been obtained 
with the VLA in its D configuration and with the Effelsberg 100-m telescope. 
A low surface density arm was discovered in the west of the galaxy, whose velocity field is distinct 
from that of the overall disk rotation. The observed gas distribution, velocity field, and velocity 
dispersion are compared to snapshots of dynamical simulations that include the effects of ram pressure.
Two different scenarios were explored: (i) ongoing stripping and (ii) a major
stripping event that took place about 300~Myr ago. It is concluded that only
the post--stripping scenario can reproduce the main observed characteristics of NGC~4569.
It is not possible to determine if the gas disk of NGC~4569 had already been truncated before
it underwent the ram pressure event that lead to its observed \HI\ deficiency.
}

\maketitle

\section{Introduction}

The spiral galaxy NGC~4569 is a very peculiar member of the Virgo cluster. Its very large
size ($D_{25}$=\am{9}{5}, or 47~kpc at the assumed Virgo cluster distance of 17~Mpc) compared to
other cluster members, and negative radial velocity ($v_{\rm r}=-235$~km\,s$^{-1}$ 
with respect to the cluster mean velocity of $\sim 1100$~km\,s$^{-1}$) have lead
to some doubt about its cluster membership, despite its close projected
distance to the cluster center ($d=1.7^{\rm o}$). Van den Bergh (1976) classified this
very red galaxy as anemic due to its low arm--interarm contrast on optical images. 
\HI\ line synthesis observations (Warmels 1988; Cayatte et al. 1990) showed that it has lost 
more than 90\% of its \HI\ gas and that its \HI\ distribution is heavily truncated.
NGC~4569 is thus an exceptional galaxy, it is one of the largest and most \HI\ deficient 
galaxies in the Virgo cluster (Solanes et al. 2001).
Since field galaxies are generally not \HI\ deficient, this is the proof that NGC~4569 is a 
Virgo cluster member whose atomic gas has been most probably stripped due to its
rapid motion through the hot intracluster medium (ICM). Nevertheless, it is not clear when
and where NGC~4569 lost its gas.

A possible hint on its stripping history may be derived from X-ray observations. 
The analysis of ROSAT observations
(Tsch\"{o}ke et al. 2001) showed that there is no soft emission (0.1--0.4~keV) from the
northern half of the disk, whereas it is pronounced in the southern disk. This is consistent with a
scenario where NGC~4569 is moving to the north-east, which causes ram pressure to push the hot 
ISM in the galaxy towards the south-west.

In addition, NGC~4569 harbors a bright, compact optical nucleus that , according to Keel (1996),
is due to a starburst dominated by supergiant stars. Tsch\"{o}ke et al. (2001)
suggest that a part of the observed extended X-ray emission is due to an outflow of hot ISM from 
the nucleus. The presence of such a feature in the west and its absence in the east is also
consistent with a galaxy motion towards the north-east. In the following we will not discuss the
possiblity of nuclear outflow, since it expands vertically to the disk and does not affect
the dynamics of the atomic gas, which is mainly located in the disk.

In the present paper we report on new deep VLA 21~cm \HI\ observations of NGC~4569, which reveal a 
new and unexpected feature - a kinetically perturbed, low surface brightness arm -
and we present models of the gas distribution and velocity field based on numerical simulations that include
the effect of ram pressure stripping. This enables us to date the latest ISM--ICM
interaction that has led to the large \HI\ deficiency of NGC~4569. The present work is
a part of a series of articles where we aim to understand the physics of ICM--ISM interactions
(Vollmer et al. 1999, Vollmer et al. 2001, Vollmer 2003, Vollmer \& Huchtmeier 2003). 
 
This article is organized in the following way: 
the observations and their results are described in Sect.~\ref{sec:observations} and
Sect.~\ref{sec:results}, respectively, the dynamical model is presented in Sect.~\ref{sec:model},
our model snapshots are compared with the observations in Sect.~\ref{sec:comparison}
and discussed in Sect.~\ref{sec:discussion}. The conclusions are given in Sect.~\ref{sec:conclusions}.

\begin{table}
      \caption{Physical parameters of NGC~4569}
         \label{tab:parameters}
      \[
         \begin{array}{lr}
            \hline
            \noalign{\smallskip}
        {\rm Other\ names} &  {\rm M~90} \\
                & {\rm VCC~1690} \\
                & {\rm UGC~7786}  \\
        $$\alpha$$\ (2000)$$^{\rm a}$$ &  12$$^{\rm h}36^{\rm m}49.8^{\rm s}$$\\
        $$\delta$$\ (2000)$$^{\rm a}$$ &  +13$$^{\rm o}09'46''$$\\
        {\rm Morphological\ type}$$^{\rm a}$$ & {\rm Sab} \\
        {\rm Distance\ to\ the\ cluster\ center}\ ($$^{\rm o}$$) & 1.7\\
        {\rm Optical\ diameter\ D}_{25}$$^{\rm a}$$\ ($$'$$) & 9.5\\
        {\rm B}$$_{T}^{0}$$$$^{\rm a}$$ & 9.79\\ 
        {\rm Systemic\ heliocentric\ velocity}$$^{\rm a}$$\ {\rm (km\,s}$$^{-1}$$)\ & -235$$\pm$$3\\
        {\rm Distance\ D\ (Mpc)} & 17 \\
        {\rm Vrot}$$_{\rm max}\ {\rm (km\,s}$$^{-1}$$) & 250$$^{\rm b}$$ \\
        {\rm PA} & 24$$^{\rm o}$$\ $$^{\rm b}$$\\
        {\rm Inclination\ angle} & 56$$^{\rm o}$$\ $$^{\rm b}$$, 62$$^{\rm o}$$\ $$^{\rm d}$$ \\
        {\rm HI\ deficiency}^{\rm c}$$ &  1.3$$\pm$$0.2\\
        \noalign{\smallskip}
        \hline
        \end{array}
      \]
\begin{list}{}{}
\item[$^{\rm{a}}$] RC3, de Vaucouleurs et al. (1991)
\item[$^{\rm{b}}$] based on H{\sc i} inematics, Guharthakurta et al. (1988)
\item[$^{\rm{c}}$] Cayatte et al. (1994)
\item[$^{\rm{d}}$] based on optical photometry, Bottinelli et al. (1983)
\end{list}
\end{table}

\section{Observations \label{sec:observations}}

\subsection{VLA}

The 21~cm line observations were made with the Very Large Array (VLA), for 
description see Napier et al. (1983). The field was centered on NGC~4569 
(see Tab.~\ref{tab:parameters}).
We observed on November 23 and 29 2001 for a total of 290 minutes with the D configuration.
The observed bandwidth of 3~MHz was devided into 63 velocity channels
of 48.8~kHz ($\sim$ 10~km\,s$^{-1}$) each. 
The velocity channels are centered on $v_{\rm hel} \sim -235$~km\,s$^{-1}$.
The data were calibrated using the standard VLA data reduction program package (AIPS).
All strong continuum sources were subtracted directly in the UV plane using a linear interpolation
of the UV data points of the first and last 10 channels.
The final rms noise per 10~km\,s$^{-1}$ wide channel at a smoothed spatial resolution of
$52'' \times 40''$ is 0.6~mJy/beam, which is close to the expected value.
We combined the UV data directly with the C array data of Cayatte et al. (1990), which were 
processed in the same way. The same weight was attributed to both data sets.
Since the C array data have a channel width of 20~km\,s$^{-1}$,
we binned every 2 channels of our D array data before joining it with the C array data.
The velocity offset between the two data sets is $\Delta v = 4$~km\,s$^{-1}$.
The resulting image of the joined UV data was CLEANed with a $39'' \times 31''$ FWHM beam.
We ended up with a r.m.s. noise of 0.5~mJy/beam in one 20~km\,s$^{-1}$ wide channel.
A flux of 1 mJy\,km\,s$^{-1}$/beam corresponds to an \HI\ column
density of $10^{18}$ cm$^{-2}$.

\subsection{Effelsberg}

On March 14-21, 2002, we performed 21-cm line observations at 6 different positions centered on
the systemic velocity of NGC~4569 with a bandwidth of 12.5~MHz. 
The two-channel receiver had a system noise of $\sim$30~K. The 1024 channel autocorrelator
was split into four banks with 256 channels each, yielding a
channel separation of $\sim$10~km\,s$^{-1}$. We further binned the channels to obtain
a final channel separation of $\sim$20~km\,s$^{-1}$ like the VLA data. 
The galaxy's central position and four 
positions at a distance of one beam width (9.3$'$) to the NW, SW, SE, and
NE from the galaxy center were observed in on--off mode
(5~min on source, 5~min off source). In addition, we observed a sixth position
\am{6}{5} west of the galaxy center. 
\begin{table}
      \caption{Integration times and rms.}
         \label{tab:table}
      \[
         \begin{array}{lcccccc}
           \hline
           \noalign{\smallskip}
           {\rm position} & {\rm C} & {\rm NW} & {\rm W} & {\rm SW} & {\rm SE} & {\rm NE} \\
	   \noalign{\smallskip}
	   \hline
	   \noalign{\smallskip}
	   $$\Delta t$$\ {\rm (min)} & 120 & 120 & 120 & 120 & 120 & 120 \\ 
           \noalign{\smallskip}
	   \hline
	   \noalign{\smallskip}
	   {\rm rms\ (mJy)} & 1.3 & 1.1 & 2.0 & 1.6 & 1.6 & 1.5 \\
	   \noalign{\smallskip}
        \hline
        \end{array}
      \]
\end{table}
Care was taken to avoid other Virgo cluster galaxies with velocities within our bandwidth
in all observations. We used 3C286 for pointing and flux calibration. 
The observation time was 120~min per position.
The resulting noise (Table~\ref{tab:table}) is partly determined by small amplitude 
interferences, but it is close to the theoretical noise of 2~mJy per hour 
of integration: on average 1.5~mJy (varying from 1.1 to 2.0~mJy).

\section{Results \label{sec:results}}

\subsection{The VLA data}

\begin{figure*}
        \resizebox{\hsize}{!}{\includegraphics{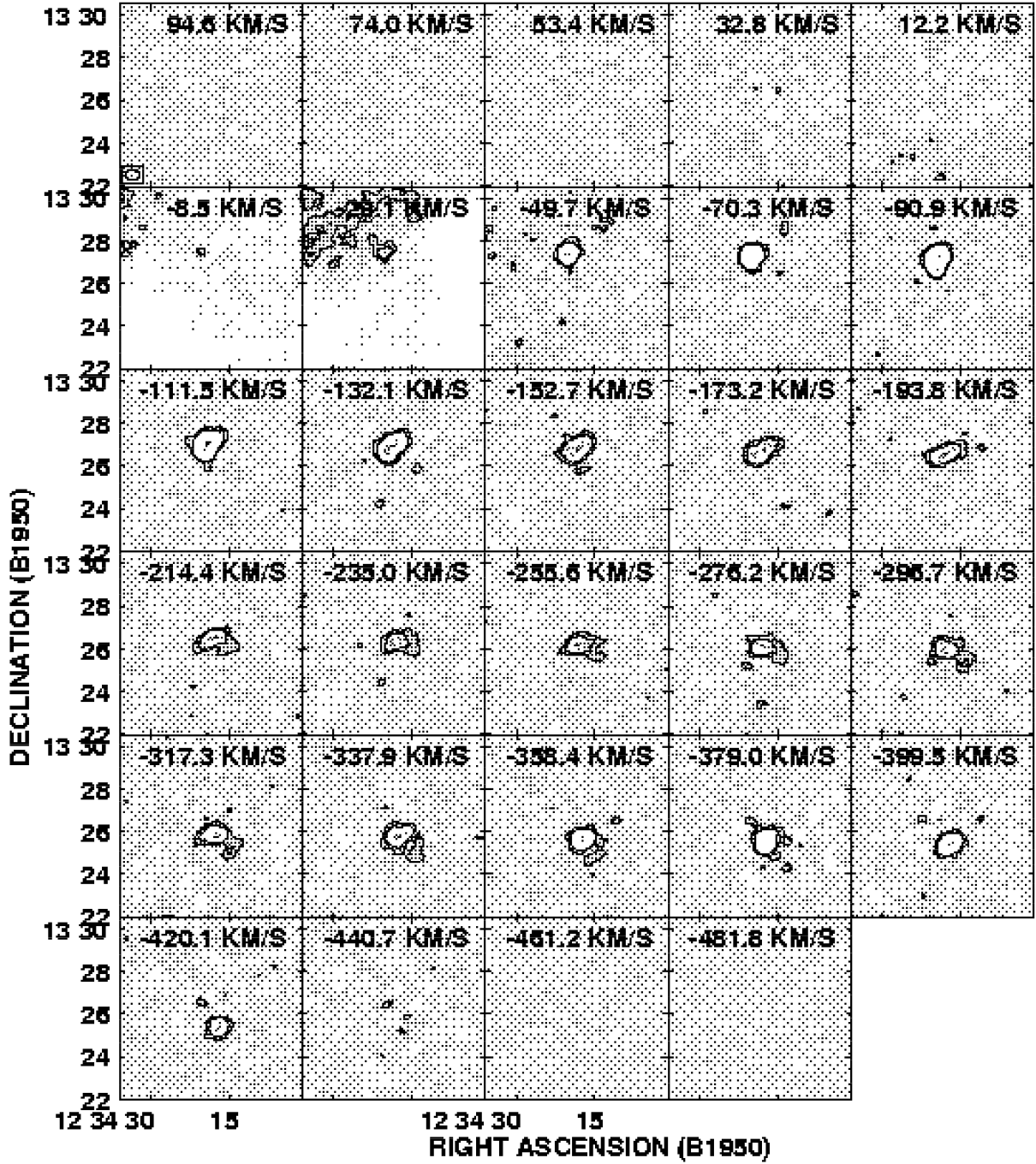}}
        \caption{VLA \HI\ channel maps of NGC~4569. The central velocity of each channel
is indicated in the upper part of each panel. The HPBW ($31''\times 39''$) is plotted in the 
lower left corner of the first channel. The contours are in steps of $3\sigma$=1.5~mJy/beam
beginning at 1.5~mJy/beam.
        } \label{fig:channels}
\end{figure*}

The channel maps are presented in Fig.~\ref{fig:channels}. Although two channels 
($v=-30$~km\,s$^{-1}$ and $-10$~km\,s$^{-1}$) are contaminated by Galactic \HI,
the \HI\ of NGC~4569 at these velocities can still be recognized and isolated from it.
The galactic rotation of NGC~4569 is clearly visible. The most interesting feature appears in the
velocity range between $-400$~km\,s$^{-1}$ and $-276$~km\,s$^{-1}$, where a detached 
emission region can be seen in the south-west part of NGC~4569, which shows rotation.

\begin{figure}
        \resizebox{\hsize}{!}{\includegraphics{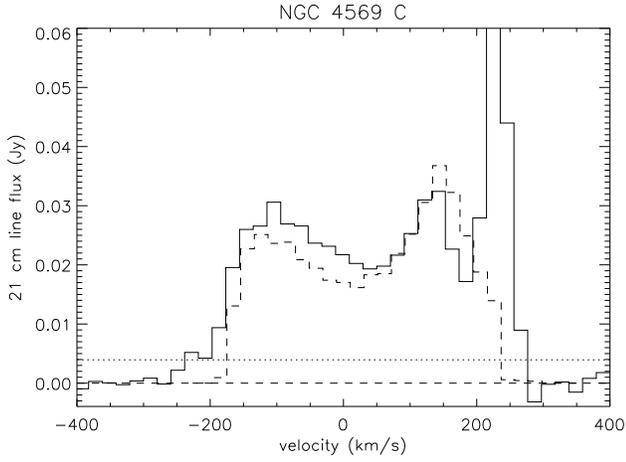}}
        \caption{Solid line: Effelsberg spectrum of the central position.
	Dashed line: integrated VLA spectrum. Dotted line: 3$\sigma$ noise level of the
	Effelsberg spectrum. Plotted are radial velocities relative to the systemic velocity
	of NGC~4569.
        } \label{fig:spectrum}
\end{figure}

We have produced an integrated \HI\ spectrum of NGC~4569 by adding all signals exceeding the 
3$\sigma$ level in all channels that contain line emission, inside an area around NGC~4569
small enough not to pick up Galactic \HI.
Fig.~\ref{fig:spectrum} shows the integrated VLA spectrum together with the Effelsberg
100-m data of the central pointing. The Effelsberg beam of $9.3'$ includes the
the whole disk \HI\ emission.
The Galactic foreground is responsible for the huge peak in the Effelsberg spectrum at 
$\sim -220$~km\,s$^{-1}$ relative to the systemic velocity, whose width 
(FWHM) is $\Delta v \sim 50$~km\,s$^{-1}$.
The total flux and peak flux density of Cayatte et al. (1990) are $\sim$20\% higher than our VLA
(C and D array)  values, 
which is within the calibration accuracy. Both of our spectra agree reasonably well, and
both peaks of the double horned shaped spectrum have about the same peak flux density.
We derive a linewidth at 20\%/50\% of the peak flux of $W_{20}=390$~km\,s$^{-1}$/
$W_{50}=350$~km\,s$^{-1}$ and a total flux of 8.85~Jy\,km\,s$^{-1}$, which corresponds to a total \HI\ 
mass of $6\,10^{8}$~M$_{\odot}$ assuming a distance of 17~Mpc.

\begin{figure}
        \resizebox{8cm}{!}{\includegraphics{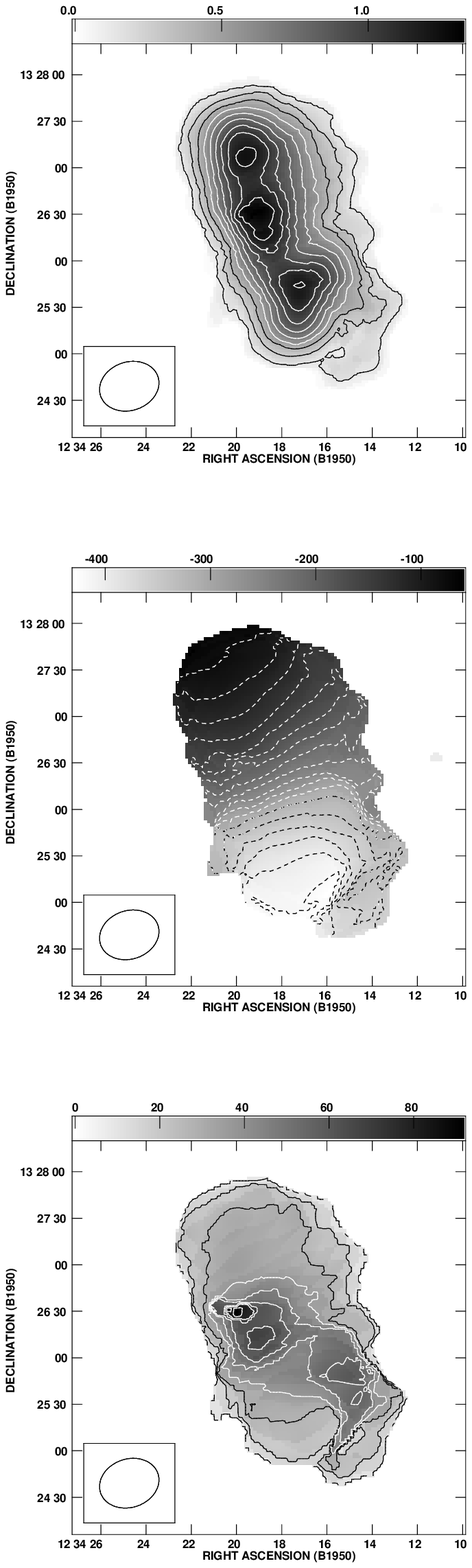}}
        \caption{\HI\ moment maps of NGC~4569. Upper panel: column density.
	  The contour steps are from 1 to $9 \times 1.3\,10^{20}$~cm$^{-2}$.
	  Middle panel: velocity field. Contour levels are from -440 to -40~km\,s$^{-1}$
	  in steps of 20~km\,s$^{-1}$. Lower panel: dispersion velocity.
	  Contour levels are from 1 to $9 \times 8.3$~km\,s$^{-1}$.
        } \label{fig:HI}
\end{figure}

\begin{figure}
        \resizebox{\hsize}{!}{\includegraphics{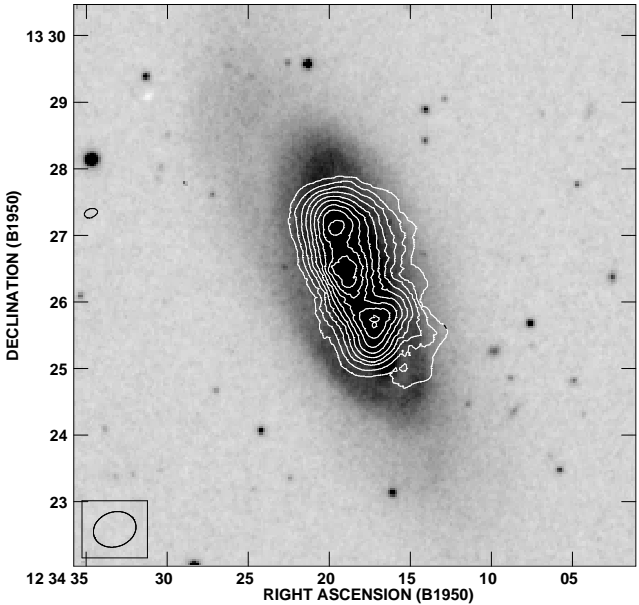}}
        \caption{Contours: \HI\ column density. The contour levels are the same as in 
	  Fig.~\ref{fig:HI}. Greyscale: B band DSS image. The contrast is chosen such that
	  the whole disk extent becomes visible.
        } \label{fig:HIopt}
\end{figure}

The first three moment maps (column density, velocity field, and dispersion velocity)
can be seen in Fig.~\ref{fig:HI}. For the determination of all moment maps the data were clipped at a
level of $3\sigma = 1.5$~mJy/beam in each channel containing \HI\ emission. 
The \HI\ distribution consists mainly of three maxima: at the center,
to the north and to the south. The southern maximum lies west from the major axis.
The \HI\ is thus not uniformly distributed within the disk, but might be distributed
along spiral arms as it is the case for NGC~4579 (Cayatte et al. 1990) or NGC~4548
(Vollmer et al. 1999). We observe a low surface density arm to the west, which was already
visible on the channel maps. This arm can also be seen on the H$\alpha$ image
presented in Tsch\"{o}ke et al. (2001) and the blue optical image shown in Fig.~\ref{fig:HIopt}.

The velocity field of NGC~4569 is quite symmetric within the disk. The roation curve is
rising on both sides with a maximum of 250~km\,s$^{-1}$ assuming the kinematic inclination 
angle of $i=56^{\circ}$ (Guharthakurta et al. 1988) or 230~km\,s$^{-1}$ assuming the inclination
based on the optical morphology $i=62^{\circ}$ (Bottinelli et al. 1983). 
The western, low surface density arm is kinematically distinct from the disk gas.
This arm shows rotation, but has systematically higher velocities than the disk gas.

As expected, the velocity dispersion shows a maximum in the galaxy center, due to
rotation and beam smearing. The absolute maximum is located $\sim 10''$ to
the north-east of the nucleus and has a velocity dispersion greater than 80~km\,s$^{-1}$, 
a second more extended maximum is observed at the inner edge of the western,
low surface density arm south of the nucleus.

\subsection{The Effelsberg data}

\begin{figure}
        \resizebox{\hsize}{!}{\includegraphics{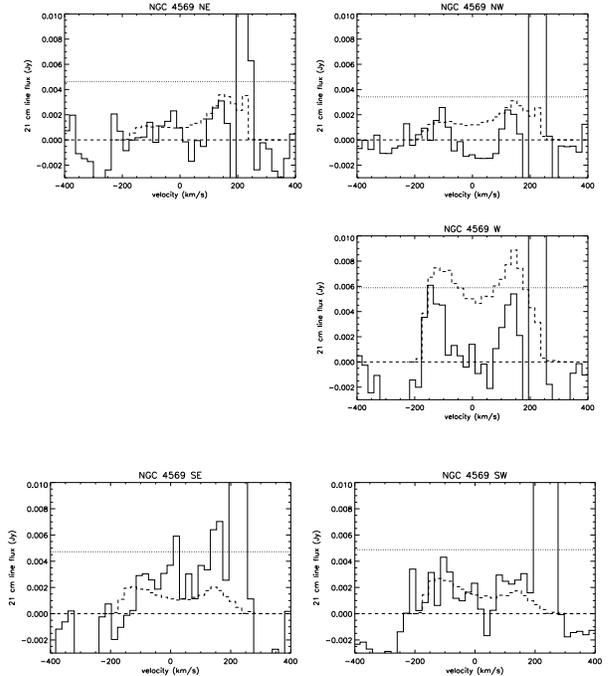}}
        \caption{Solid lines: Effelsberg 100-m spectra of the five off-center positions.
	Their locations with respect to the galaxy center are marked on top of each panel.
	Dashed line: synthesized VLA spectra, which only show \HI\ disk emission. 
	Dotted line: 3$\sigma$ noise levels of the Effelsberg spectra. 
	Radial velocities are given relative to the systemic velocity of NGC~4569.
        } \label{fig:spectra}
\end{figure}

For the 4 off-center Effelsberg positions we have synthesized spectra from the VLA data 
using the area defined by the pointing positions and the (\am{9}{3}) Effelsberg HPBW.
In this way we take advantage of the spatial \HI\ distribution of the VLA data
(see Vollmer \& Huchtmeier 2003). 
Fig.~\ref{fig:spectra} shows the Effelsberg spectra of these positions (solid lines)
together with the synthesized VLA spectra. Since the VLA data cube shows only \HI\ gas
located in the disk (see Cayatte et al. 1990), any emission excess
in the off-center Effelsberg spectra must be due to gas located outside the disk.
No such excess is seen and an upper mass limit $\sim 10^{7}$~M$_{\odot}$
can be set for the \HI\ outside the disk, corresponding to a signal of $5 \sigma$ in a 20~km\,s$^{-1}$ 
channel with the an rms of $\sigma=1.6$~mJy.
The absence of atomic gas far away from the galaxy ($>20$~kpc), as was found for
NGC~4388 (Vollmer \& Huchtmeier 2003), is an important constraint on dynamic
stripping models.

\section{Simulations \label{sec:model}}

We have adopted a model where the ISM is simulated as a collisional component,
i.e. as discrete particles which possess a mass and a radius and which
can have inelastic collisions (sticky paticles).
Since the ISM is a turbulent and fractal medium (see e.g. Elmegreen \& Falgarone 1996),
it is neither continuous nor discrete. The volume filling factor of the different phases
is smaller than one. The warm neutral and ionized gas fill about $50\%$ of the volume,
whereas cold neutral gas has a volume filling factor of only 2\% (Boulares \& Cox 1990). 
It is not clear how this fraction changes, when an external 
pressure is applied. In contrast to smoothed particles hydrodynamics (SPH), which is a 
quasi continuous approach and where the particles cannot penetrate each other, our approach 
allows a finite penetration length, which is given by the mass-radius relation of the particles.
Both methods have their advantages and their limits.
The advantage of our approach is that ram pressure can be included easily as an additional
acceleration on particles that are not protected by other particles (see Vollmer et al. 2001).
In this way we avoid the problem of treating the huge density contrast between the 
ICM ($n \sim 10^{-4}$~cm$^{-3}$) and the ISM ($n > 1$~cm$^{-3}$) of the galaxy.

Since the model is described in detail in Vollmer et al. (2001), we 
will summarize only its main features.
The N-body code consists of two components: a non-collisional component
that simulates the stellar bulge/disk and the dark halo, and a
collisional component that simulates the ISM.
The 20\,000 particles of the collisional component represent gas cloud complexes which are 
evolving in the gravitational potential of the galaxy.

The total assumed gas mass is $M_{\rm gas}^{\rm tot}=6.2\,10^{9}$~M$_{\odot}$,
which corresponds to the total neutral gas mass before stripping, i.e.
to an \HI\ deficiency of 1.0, which is defined as the logarithm of the ratio between
the \HI\ content of a field galaxy of same morphological type and diameter
and the observed \HI\ mass.
To each particle a radius is attributed depending on its mass. 
During the disk evolution the particles can have inelastic collisions, 
the outcome of which (coalescence, mass exchange, or fragmentation) 
is simplified following Wiegel (1994). 
This results in an effective gas viscosity in the disk. 

As the galaxy moves through the ICM, its clouds are accelerated by
ram pressure. Within the galaxy's inertial system its clouds
are exposed to a wind coming from a direction opposite to that of the galaxy's 
motion through the ICM. 
The temporal ram pressure profile has the form of a Lorentzian,
which is realistic for galaxies on highly eccentric orbits within the
Virgo cluster (Vollmer et al. 2001).
The effect of ram pressure on the clouds is simulated by an additional
force on the clouds in the wind direction. Only clouds which
are not protected by other clouds against the wind are affected.

The non--collisional component consists of 49\,125 particles, which simulate
the galactic halo, bulge, and disk.
The characteristics of the different galactic components are shown in
Table~\ref{tab:param}.
\begin{table}
      \caption{Total mass, number of particles $N$, particle mass $M$, and smoothing
        length $l$ for the different galactic components.}
         \label{tab:param}
      \[
         \begin{array}{lllll}
           \hline
           \noalign{\smallskip}
           {\rm component} & M_{\rm tot}\ ({\rm M}$$_{\odot}$$)& N & M\ ({\rm M}$$_{\odot}$$) & l\ ({\rm pc}) \\
           \hline
           {\rm halo} & 2.4\,10$$^{11}$$ & 16384 & $$1.4\,10^{7}$$ & 1200 \\
           {\rm bulge} & 8.2\,10$$^{9}$$ & 16384 & $$5.0\,10^{5}$$ & 180 \\
           {\rm disk} & 4.1\,10$$^{10}$$ & 16384 & $$2.5\,10^{6}$$ & 240 \\
           \noalign{\smallskip}
        \hline
        \end{array}
      \]
\end{table}
The resulting rotation velocity is $\sim$220~km\,s$^{-1}$ and the rotation curve
is flat. 

The particle trajectories are integrated using an adaptive timestep for
each particle. This method is described in Springel et al. (2001).
The following criterion for an individual timestep is applied:
\begin{equation}
\Delta t_{\rm i} = \frac{20~{\rm km\,s}^{-1}}{a_{\rm i}}\ ,
\end{equation}
where $a_{i}$ is the acceleration of the particle i.
The minimum value of $t_{\rm i}$ defines the global timestep used 
for the Burlisch--Stoer integrator that integrates the collisional
component.

The galaxy is on an eccentric orbit within the cluster. The temporal
ram pressure profile can be described by:
\begin{equation}
p_{\rm ram}=\frac{p_{\rm max}}{t^{2}+t_{\rm HW}^{2}}\ ,
\end{equation}
where $t_{\rm HW}$ is the width of the profile (Vollmer et al. 2001). 
We set $p_{\rm max}$=4000~cm$^{-3}$(km/s)$^{2}$ and $t_{\rm HW}$=50~Myr.
The efficiency of ram pressure also depends on the inclination angle $i$
between the galactic disk and the orbital plane (Vollmer et al. 2001).
We set $i$=35$^{\rm o}$. The model parameters are listed in Table~\ref{tab:modelparameters}.
\begin{table}
      \caption{Model parameters for time dependent ram pressure stripping}
         \label{tab:modelparameters}
      \[
         \begin{array}{lr}
            \hline
            \noalign{\smallskip}
        {\rm maximum\ ram\ pressure}\ \big({\rm cm}$$^{-3}$$({\rm km\,s}$$^{-1}$$)$$^{2}$$\big)\ \ &  4000\\
        {\rm inclination\ angle\ between\ orbital\ and\ disk\ plane} & 35$$^{\rm o}$$ \\
        {\rm final\ HI\ deficiency} &  1.0\\
        \noalign{\smallskip}
        \hline
        \end{array}
      \]
\end{table}

\begin{figure*}
        \resizebox{\hsize}{!}{\includegraphics{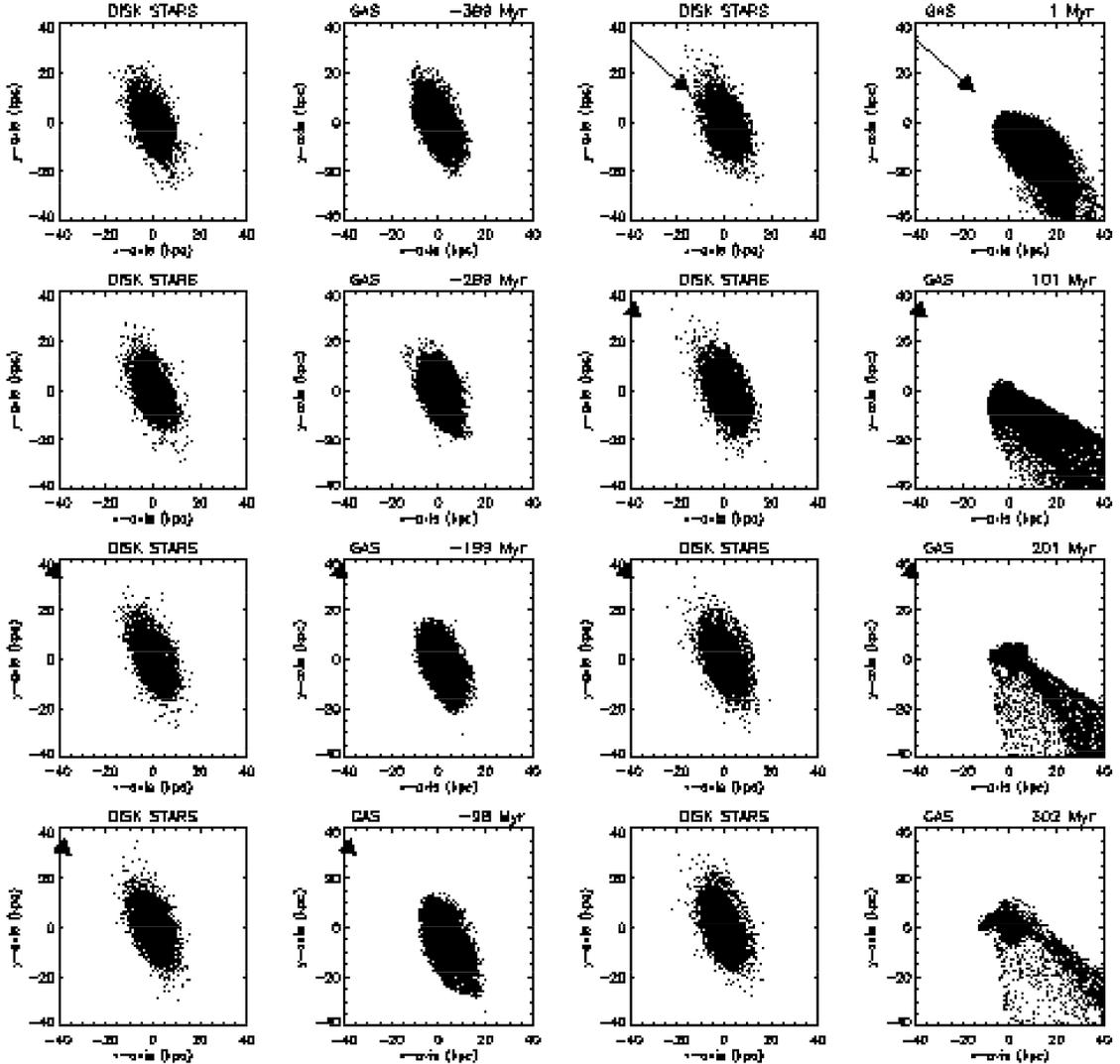}}
        \caption{Evolution of the model stellar (1st and 3rd column)
	and gas disk (2nd and 4th column). 
	The major axis position angle and inclination of NGC~4569 
	are $PA=23^{\rm o}$ and $i=62^{\rm o}$, respectively. The arrow
	indicates the direction of ram pressure, i.e. it is opposite to
	the galaxy's velocity vector, and its size is proportional
	to $\rho v_{\rm gal}^{2}$. The galaxy passes the cluster core at
	0~Myr. The time to the core passage is marked at the top of each panel.
        } \label{fig:evolution}
\end{figure*}

The large scale evolution of NGC~4569 can be seen in Fig.~\ref{fig:evolution}.
As expected, the stellar disk does not change, because ram pressure selectively
affects only the gas. During the passage of NGC~4569 through the cluster core
($t=0$~Myr corresponds to the minimum distance between the galaxy and M87)
its gas is pushed strongly towards the south-west and finally stripped at 
$t > 50$~Myr. The gas that was pushed away, but not accelerated to
the escape velocity, falls back onto the galaxy at $t > 150$~Myr. This re-accretion
of the gas has to be regarded with caution, since its occurence depends on the gas physics
(cloud expansion, evaporation, ISM--ICM phase mixing) used in the model. 
We assume that the column density of the clouds
does not change during the interaction -- however, if it were to decrease due to, e.g., evaporation
or ICM--ISM mixing, then re-accretion will be suppressed. 

\begin{figure}
        \resizebox{\hsize}{!}{\includegraphics{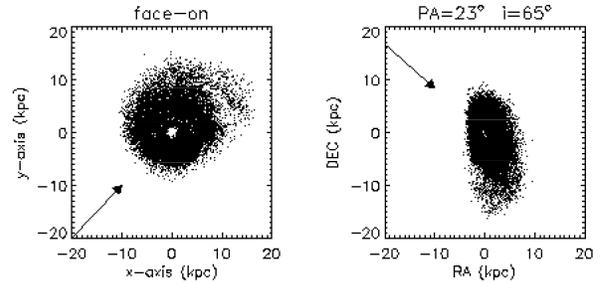}}
        \caption{Comparison snapshot of the gas distribution of NGC~4569 for
	  the case of ongoing ram pressure stripping. Left panel: the disk 
	  seen face-on. Right panel: view of the disk using the observed position angle and 
	  inclination of NGC~4569. The arrow indicates the direction of the wind,
	  i.e. opposite to the galaxy's motion.
        } \label{fig:compon}
\end{figure}

During the whole galaxy evolution within the cluster when ram pressure is important
only two scenarios can give rise to an asymmetric gas arm:
(i) during the resettlement of the gas after a strong stripping event and
(ii) during the phase of ongoing stripping.
Therefore, we compare two snapshots with our \HI\ observations: 
(i) the last snapshot of Fig.~\ref{fig:evolution} ($t=300$~Myr), i.e. a past stripping event 
and (ii) the case of ongoing
stripping -- another simulation made using a truncated gas disk at $R=10$~kpc.
with $p_{\rm max}$=2000~cm$^{-3}$(km/s)$^{2}$, $i=35^{\rm o}$, 
$t_{\rm HW}$=80~Myr. The comparison snapshot at $t=0$~Myr is shown in Fig.~\ref{fig:compon}.
For a  given position angle and inclination, the third angle, which is the azimuthal
angle in the plane of the galaxy, has to be fixed. A variation of the azimuthal angle
leads to a variation of the 3D direction of the galaxy's motion with respect to the observer.
We chose the value that reproduces best the observed, western, low surface density arm.
In both cases the galaxy moves to the north-east. We made sure that the galaxy's radial 
velocity did not fall below $-1500$km\,s$^{-1}$. 

We constructed model data cubes of the two snapshots and made moment maps with AIPS.
Only gas particles with a local gas density greater than 100~cm$^{-3}$ where 
put into the model cube.

\section{Comparison between simulations and observations \label{sec:comparison}}

\begin{figure}
        \resizebox{7cm}{!}{\includegraphics{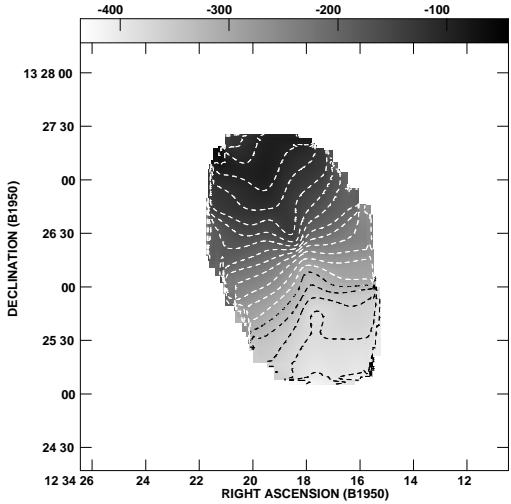}}
        \caption{Initial gas velocity field. The disk is truncated at a radius of 5~kpc.
        } \label{fig:initvel}
\end{figure}

In this Section we compare model spectra and the three model moment maps (gas distribution, velocity field,
and dispersion velocity) to those of our \HI\ observations (Fig.~\ref{fig:spectrum}, \ref{fig:HI},
and \ref{fig:spectra}). The initial velocity field of the gas disk is shown in Fig.~\ref{fig:initvel}.
To facilitate the comparison with the final snapshots of our simulation, we have truncated the disk 
at a radius of 5~kpc. Our model velocity field has a steeper rotation curve and reaches
the plateau of constant rotation velocity at a smaller radius than the observed one (Fig.~\ref{fig:HI}).
However, the CO position-velocity diagram along the major axis of NGC~4569 
(Fig.~23 of Kenney \& Young 1988) shows a steeper rotation curve, which is close to that of our model.

\subsection{Ongoing stripping}

\begin{figure}
        \resizebox{8cm}{!}{\includegraphics{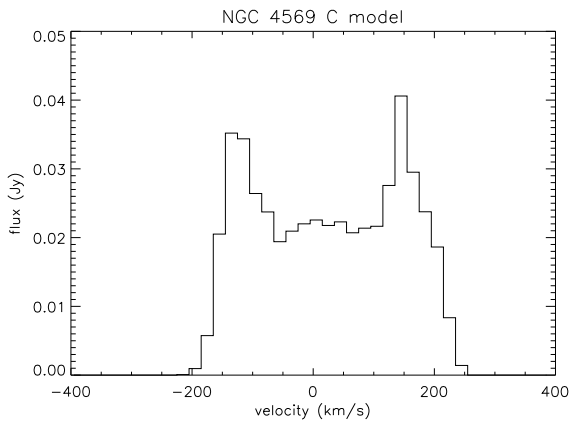}}
        \caption{Model \HI\ spectrum for ongoing ram pressure stripping.	  
        } \label{fig:cspectrum_139}
\end{figure}

\begin{figure}
        \resizebox{8cm}{!}{\includegraphics{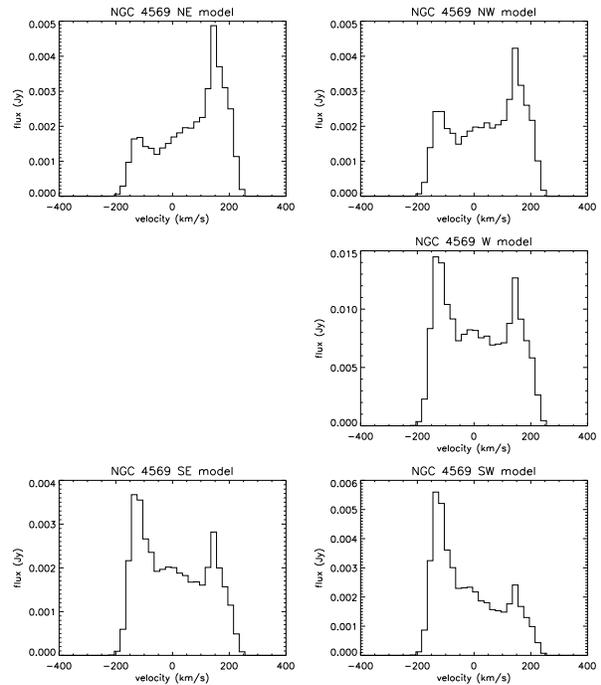}}
        \caption{Model spectra corresponding to our offset \HI\ observations (Fig.~\ref{fig:spectra}) 
	  for ongoing stripping.
        } \label{fig:spectra_139}
\end{figure}

The model spectra corresponding to the observed center spectrum and those of the four
offset \HI\ observations (Fig.~\ref{fig:spectra}) for ongoing stripping are shown
in Fig.~\ref{fig:cspectrum_139} and \ref{fig:spectra_139}. 
The two peaks of the central double-horn spectrum have about the same velocity width and the
edges of the profile are sharp. The off-center spectra in the NE, SE, and NW show maximum
flux densities smaller than the detection limit of our Effelsberg data ($\sim 5$~mJy).
The signature of the western arm is clearly visible at negative velocities in the
W and SW spectra.

\begin{figure}
        \resizebox{8cm}{!}{\includegraphics{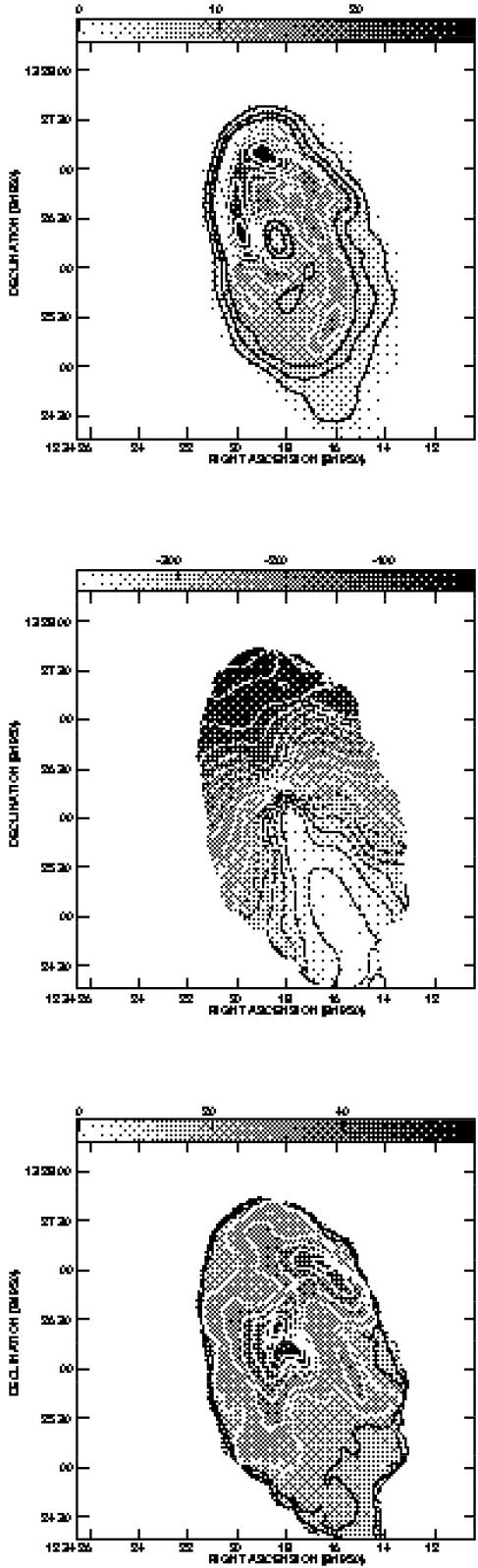}}
        \caption{Model moment maps for ongoing ram pressure stripping.
	  Upper panel: gas distribution in arbitrary units. The relative contours 
	  are the same as in Fig.~\ref{fig:HI}. Middle panel: velocity field.
	  The contours are the same as in Fig.~\ref{fig:HI}. Lower panel:
	  velocity dispersion. The contours are the same as in Fig.~\ref{fig:HI}.	  
        } \label{fig:HI_acmodel}
\end{figure}

Fig.~\ref{fig:HI_acmodel} shows the moment maps for ongoing ram pressure stripping.
As already mentioned, the projection parameters are chosen such that (i) the radial
velocity of NGC~4569 and (ii) the observed, western, low surface density arm are reproduced.
The central hole in the gas distribution is artificial, in order to avoid too small timesteps
for the particle integration. The north-eastern maximum represents compressed gas that has been
pushed by ram pressure to smaller galactic radii.
The velocity field in the north is that of a rising rotation curve, because the ISM
is pushed inwards by the ICM. This effect is also observed for NGC~4654 (Vollmer 2003).
The southern part of the disk velocity field shows a plateau of constant rotation velocity. 
The velocity contours are continuous from the disk to the western, low surface density arm,
because the gas there is accelerated and pushed out of the galaxy in the sense of rotation.
As expected, the maximum of the velocity dispersion is located in the galaxy center.
A second maximum can be found in the north-west. It corresponds to gas that has been
pushed inwards in the north-east by the ICM and has rotated (the galaxy is rotating clock-wise) 
towards the west.

\subsection{Past stripping event} 

\begin{figure}
        \resizebox{8cm}{!}{\includegraphics{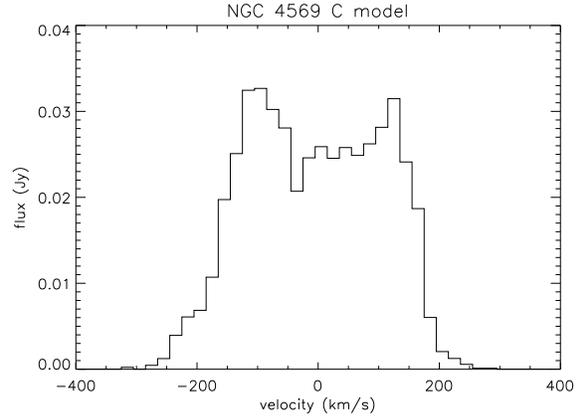}}
        \caption{Model \HI\ spectrum for a past ram pressure stripping event.	  
        } \label{fig:cspectrum_160}
\end{figure}

\begin{figure}
        \resizebox{8cm}{!}{\includegraphics{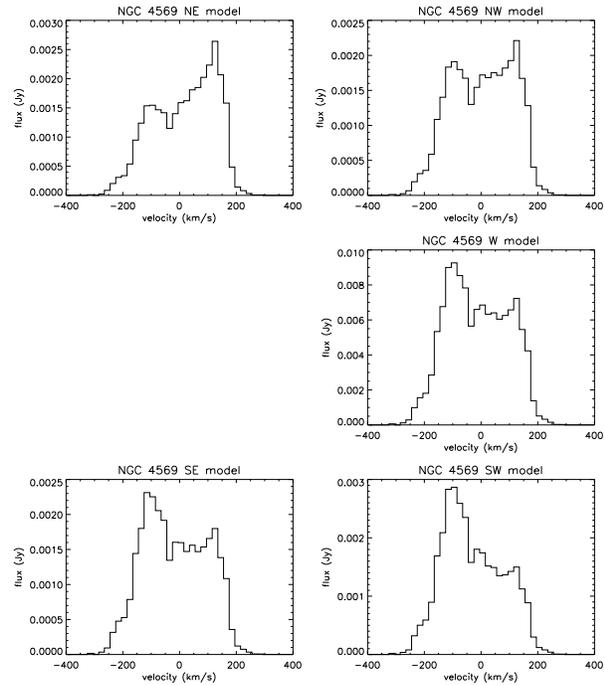}}
        \caption{Model spectra corresponding to our offset \HI\ observations (Fig.~\ref{fig:spectra}) 
	  for a past ram pressure stripping event.
        } \label{fig:spectra_160}
\end{figure}

The model spectra corresponding to the observed center spectrum and those of the four
offset \HI\ observations (Fig.~\ref{fig:spectra}) for a past stripping event are shown
in Fig.~\ref{fig:cspectrum_160} and \ref{fig:spectra_160}. 
The low velocity peak of the central double-horn spectrum has a larger velocity width 
than the high velocity peak. Moreover, the low-velocity edge of the profile shows a pronounced
wing. All off-center spectra except the one in the W show maximum
flux densities smaller than the detection limit of our Effelsberg data ($\sim 5$~mJy).

\begin{figure}
        \resizebox{8cm}{!}{\includegraphics{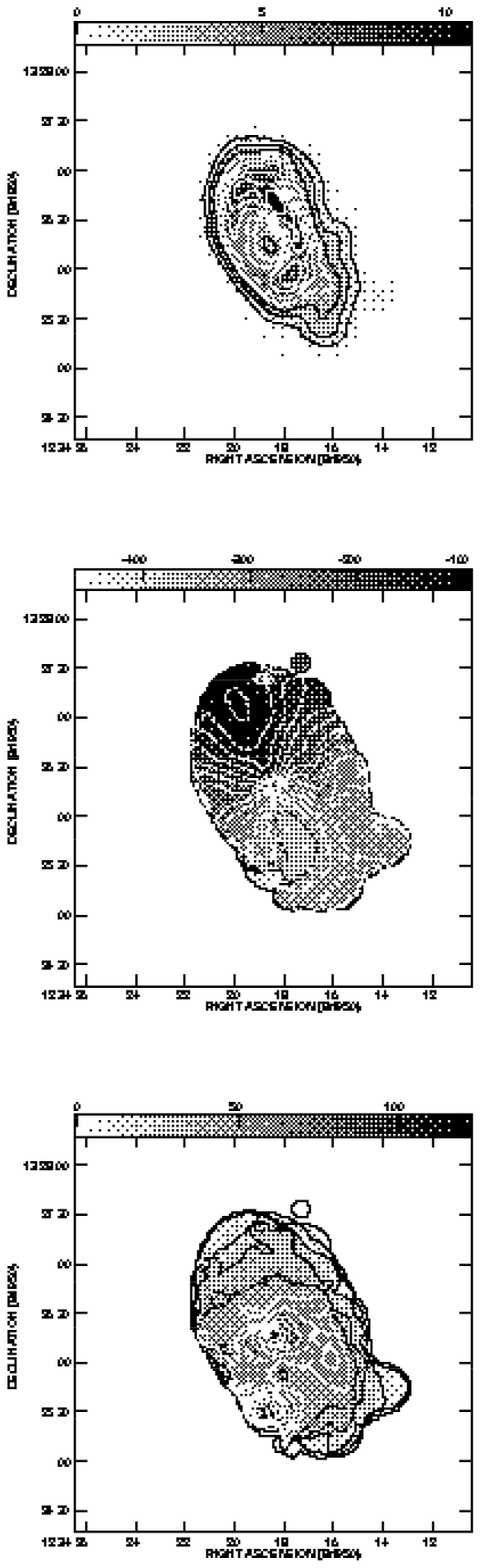}}
        \caption{Model moment maps for a past ram pressure stripping event.
	  Upper panel: gas distribution in arbitrary units. The relative contours 
	  are the same as in Fig.~\ref{fig:HI}. Middle panel: velocity field.
	  The contours are the same as in Fig.~\ref{fig:HI}. Lower panel:
	  velocity dispersion. The contours are the same as in Fig.~\ref{fig:HI}.
        } \label{fig:HI_model}
\end{figure}

Fig.~\ref{fig:HI_model} shows the moment maps for a past ram pressure stripping event.
Again, the projection parameters are chosen such that they reproduce (i) the radial
velocity of NGC~4569 and (ii) the observed, western, low surface density arm.
Again, the central hole in the gas distribution is artificial, in order to avoid too small timesteps
for the particle integration. The spatial extent of the model gas distribution is
smaller than that of the observed one, because we have decided to reproduce the \HI\ deficiency
rather than the spatial extent. 
For simplicity we assume that all gas in the model is atomic. Consequently, our inital central gas surface density 
is higher than the observed one (our model rather represents an Sc/Sb galaxy, whereas NGC~4569
is classified as Sab). Thus, for a given \HI\ deficiency, our model gas disk is smaller 
than the observed disk.
The disk velocity field shows a plateau on both sides, which is reached at smaller
galactic radii in the south than in the north. The velocity field of the southern part
of the western, low surface density arm is not a continuation of that of the disk.
The absolute velocities of the arm are higher than those of the neighbouring disk.
The maximum of the velocity dispersion is located in the center. A second, point-like
maximum can be found to the south-east of the nucleus where the radial velocity
has its minimum value. This feature is due to a gas streamer that falls onto the
galaxy from behind. In addition, we observe an 
extended maximum at the inner, edge of the western, low surface density arm south of the nucleus.

\section{Discussion \label{sec:discussion}}

Since both models reproduce the observed, western, low surface density arm as well as the
radial velocity of NGC~4569, it became clear that it is impossible to
chose between the two scenarios only on the basis of the observed gas distribution.
Only the velocity information in addition to the gas distribution allows us
to discriminate.

The extended low velocity edge of the observed central \HI\ spectrum is only
reproduced by the past stripping simulation. In general, the offset spectra of both
simulations are consistent with our observations, i.e. only in the west (W)
atomic gas is detected (flux density greater than $\sim 5$~mJy in one
20~km\,s$^{-1}$ channel). The western arm of the ongoing stripping model is somewhat too
prominent in the model SW spectrum, but this is due to the detailed choice of the stripping
parameters. It is important to note that the tail of stripped and back-falling gas of the post-stripping 
model (Fig.~\ref{fig:evolution}) could not have been detected with the Effelberg telescope
(Fig.~\ref{fig:spectra} neither with the VLA).

Ongoing stripping steepens the velocity field in the direction of the galaxy's motion
and elongates the velocity contours on the opposite side (Fig.~\ref{fig:HI_acmodel}; 
see also Vollmer 2003), because ram pressure pushes the gas to smaller galactic radii in the 
direction of the galaxy's motion and accelerates it in the sense of rotation on the other side.
On the other hand, a past stripping event causes the steepening and asymmetry of the rotation curve 
(Fig.~\ref{fig:HI_model}). This asymmetry is also seen
in the observed velocity field (Fig.~\ref{fig:HI}) where the rotation curve of the souther half is 
steeper than that in the north. Regarding the general steepening of the
rotation curve: since we do not know the initial 
velocity field of NGC~4569, we do not think that the difference between the simulated 
(Fig.~\ref{fig:HI_model}) and observed (Fig.~\ref{fig:HI}) velocity fields poses a problem
for the validity of the simulations.

Whereas the velocity field in the western arm region is continuous in the case
of ongoing stripping, it shows a discontinuity in the post-stripping scenario
as is observed. Also, the difference in absolute velocity between the disk and
the western arm of the post-stripping scenario fits nicely our \HI\ observations.
As for the distribution of dispersion velocities, the observed, single, local, extended 
maximum at the inner edge of the western am is only reproduced by the post-stripping 
model. It is located at the interface between the western, low surface density
arm and the gas rotating within the galactic disk. We failed to find any suspect 
features in the 3D gas distribution at that location.
The general dynamics are very complicated, which makes it hard to disentangle
different physical effects due re-accretion, gas expansion, and rotation.   
The lines in this region are broadened to both larger and smaller radial velocities compared 
to a region south of it. We have no clear explanation why this region shows an enhanced 
dispersion velocity.

There is also a point-like maximum visible, like the one observed,
but it is located to the south. The enhanced linewidth arises from
a back-falling gas streamer. Since the exact dynamics of such a streamer depends on
gas physics (e.g. volume filling factor, shock heating, evaporation) and the detailed 
stripping parameters, we do not expect to fit its exact location.
We can only speculate that such a region of unusual large velocity dispersion might be
a sign of re-accretion. 

We thus conclude that the ongoing stripping
scenario is ruled out based on the comparison of the observed and simulated velocity
fields and velocity dispersion distributions. On the other hand, the post-stripping 
scenario reproduces the major characteristics of the observations:
the truncated gas disk, the western, low surface density arm, and the discontinuity
between the velocity field in the galactic disc and the western arm.
In order to investigate if the gas disk of NGC~4569 was already truncated before it 
underwent the ram pressure stripping event that lead to the observed \HI\ deficiency, we 
made an additional simulation starting with a truncated gas disk at $R=10$~kpc.
The resulting moment maps are not distinguishable from that of Fig.~\ref{fig:HI_model}.
Thus we conclude that it is not excluded that NGC~4569 has already experienced a
ram pressure event more than $\sim 1$~Gyr ago that lead to a truncated \HI\ disk.
After such a first ram pressure stripping event the gas disk had enough time to
evolve into an equilibrium state without major perturbations. Possibly, the gas disk
became more robust via the 'disk annealing process' described by Schulz \& Struck (2001).
This is consistent with our finding that the rotation curve steepens after the
ram pressure event.
During the galaxy's way through the outer parts of the cluster some of its ISM
might have been replenished by the released gas by dying stars. All we can say is that 
NGC~4569 had a less truncated gas disk before the last ram pressure stripping event than it has now.
For each stripping event, the following parameters are important for the efficiency of
ram pressure stripping: the gas disk truncation radius, the galaxy orbit, and the inclination 
angle between the galaxy's disk and the orbital plane during the cluster core passage.
Different core passages can have different maximum ram pressures (Vollmer et al. 2001),
and the one with the highest maximum ram pressure is decisive for the galaxy.
NGC~4569 experienced its most efficient ram pressure stripping event $\sim 300$~Myr ago.

\begin{figure}
        \resizebox{\hsize}{!}{\includegraphics{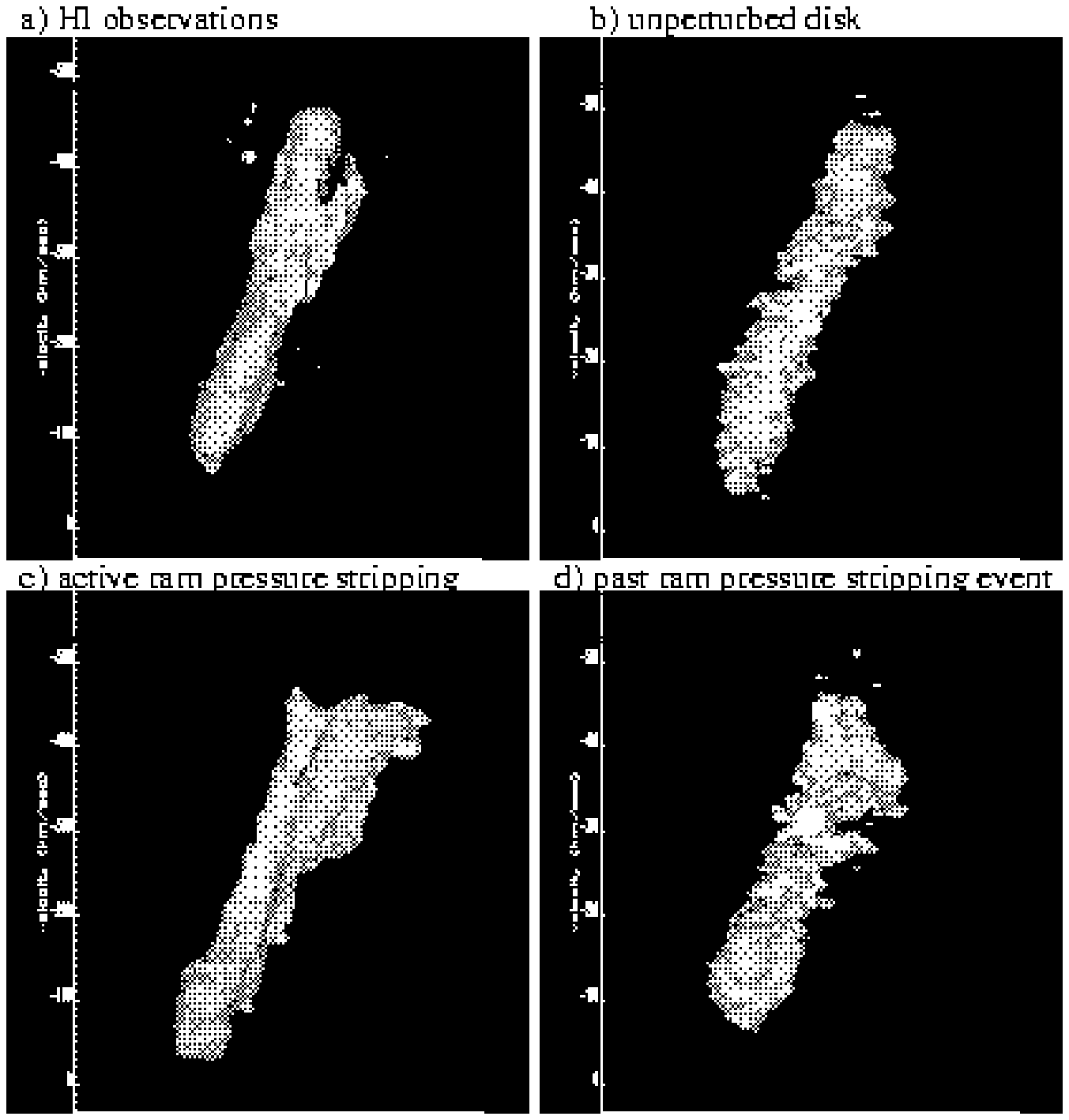}}
        \caption{Three dimensional view of the data cubes. The RA--Dec plane
	  is perpendicular to the image plane, and the velocity axis is the
	  vertical axis. a) \HI\ observations. b) Unperturbed gas disk. c)
	  Ongoing ram pressure stripping. d) Past ram pressure stripping event.
        } \label{fig:3D}
\end{figure}

A final comparison of the data cubes can be directly made in 3 dimensional representations.
We have developed a 3D visualisation of data cubes which enables us to easily
distinguish different kinematic features (see Vollmer et al. 1999). 
All velocity channels seen in Fig.~\ref{fig:channels}
are piled up to give the cube, whose axes are the right ascension, 
the declination, and the radial velocity. All points in the cube having 
intensities exceeding a chosen level (2~mJy/beam) become opaque, the rest
remain transparent. The surface created in this way is illuminated
by light which is coming out of the observer's direction,
making brighter features look closer to the observer.
If we had, for example, a sphere of increasing intensity to the centre,
it would appear as a ball with its radius increasing with decreasing
chosen intensity level. This representation allows to analyse the whole cube from any
possible point of view. In Fig.~\ref{fig:3D} we have rotated the cube
in a way that the RA--Dec plane is perpendicular to the image plane, and the velocity axis is the
vertical axis. The kinematics of rotation form a disk-like structure in the
restricted phase space (RA, DEC, $v_{\rm r}$). We have rotated the cube such that
this disk structure is seen edge-on. 
The observed, western, low surface density arm can be seen in the upper right part of
the image and is clearly distinct from the galaxy rotation (Fig.~\ref{fig:3D}a).
For comparison we show the unperturbed gas disk in Fig.~\ref{fig:3D}b.
The western arm in the ongoing stripping scenario (Fig.~\ref{fig:3D}c)
is spatially more extended and extends to smaller velocities. That in the post-stripping
scenario (Fig.~\ref{fig:3D}d) nicely fits the 3 dimensional extent of the
observed western arm. However, due to the southern plateau of the model velocity field,
the southern part of the disk-like structure in the restricted phase space is thicker
than the observed one.  We thus conclude that the post-stripping scenario is the most
probable one for NGC~4569.

This galaxy is not the first one, where an extended, low surface density arm 
has been discovered in deep \HI\ imaging observations, as an
extended, low surface density arm has also been found in three other Virgo spiral galaxies: 
NGC~4321 (Knapen et al. 1993),
NGC~4654 (Phookun \& Mundy 1995) and NGC~4548 (Vollmer et al. 1999). An extreme case
is NGC~4388 where Vollmer \& Huchtmeier (2003) detected neutral hydrogen more than
20~kpc above the disk plane. These low surface density features stand in contrast to
high surface brightness, extraplanar emission regions, which are located close to the disk.
These have been observed in NGC~4522 (Kenney et al. 2003)
and NGC~4438 (Cayatte et al. 1990), two galaxies that are most probably actively stripped.
 
The question arises if the high and low surface density features are caused by two different 
stages of an ICM--ISM interaction.
We might speculate that the low surface brightness features whose velocity fields are not a
continuation of that of the neighbouring disk are reminiscent of a past stripping event
(NGC~4569 and NGC~4548), whereas those with a continuous velocity field are due to weak
ongoing ram pressure stripping, perhaps in addition to a gravitational interaction
(NGC~4654). On the other hand, high surface brightness features appear to be evidence of a major
ongoing stripping process.

If we assume a temporal ram pressure profile as presented in Vollmer et al. (2001) 
the observational groups might be understood in the following way:

(i) strong stripping leading to a large \HI\ deficiency (NGC~4388, NGC~4438, NGC~4522,
NGC~4548, NGC~4569): during the maximum ram pressure phase high surface density gas is
pushed out of the galactic disk (NGC~4438, NGC~4522). When the pushed-out gas moves to
larger distances from the galaxy it expands and evaporates, leading to low surface
density \HI\ (NGC~4388). During active ram pressure stripping high surface density gas is 
accelerated and pushed to smaller galactic radii in the direction of the galaxy's motion and to 
larger galactic radii at the opposite direction. When the galaxy leaves the cluster core,
the ram pressure decreases strongly and the gas is left with just the gravitational potential 
of the galaxy, to which it tries to adapt itself. The compressed region expands and the
pushed-away gas falls back to the galaxy. This creates expanding, asymmetric ring structures
whose surface density decreases during their expansion. 
These short-lived ($<10^{8}$~yr) features represent the kinetically perturbed low surface density arms 
observed in NGC4569 (and maybe NGC~4321 and NGC4548, whose cases are less clear).
The number of observed low surface density arms in Virgo cluster spirals is small, 
but since these observations are time consuming, only few galaxies have been observed deep
enough to allow their detection.
However, all observations of sufficient depth have shown the presence of a low surface density 
arm.

(ii) weak stripping together with a gravitational interaction leading to a small or even
negligible \HI\ deficiency: ram pressure pushes away the low surface density gas which has 
been torn out of the galactic disk by the gravitational interaction (NGC~4654).

This hypothesis needs to be verified with a larger sample of cluster spirals.

\section{Conclusions \label{sec:conclusions}}

We obtained deep \HI\ line observations of the anemic Virgo spiral galaxy
NGC~4569 with the VLA in its D configuration and with the Effelsberg 100-m telescope.
Snapshots of a dynamical model including the effects of ram pressure 
were compared to the observations. Two different stages of a ram pressure stripping event
were investigated: (i) ongoing stripping and (ii) a past stripping event.
The results of our investigations are:
\begin{enumerate}
\item
we detected a low surface density arm in the west of the galaxy;
\item
the velocity field of this arm is not a continuation of the velocity field of disk
rotation;
\item
the absolute radial velocities of the arm are higher than those of the neighbouring disk;
\item
the velocity dispersion shows a local maximum at the inner edge of the arm;
\item
there is no gas mass greater than $10^{7}$~M$_{\odot}$ detected far away ($>20$~kpc) from the galaxy;
\item
the post-stripping scenario can reproduce the main characteristics of the
gas distribution, velocity field, and velocity dispersion together with
the radial velocity of the galaxy;
\item
within this scenario the galaxy has passed through the cluster core about 300~Myr ago;
\item
it is not excluded that NGC~4569 had an already truncated gas disk before this ICM--ISM
interaction.
\end{enumerate}

\begin{acknowledgements}
Based on observations with the 100-m telescope of the MPIfR (Max-Planck-Institut f\"{u}r 
Radioastronomie) at Effelsberg. The Very Large Array of the National Radio Astronomy
Observatory (NRAO) is operated by Associated Universities Inc., under contract with the
National Science Foundation.
\end{acknowledgements}


\begin{thebibliography}{}

\bibitem{a1} Bottinelli L., Gouguenheim L., \& Paturel G. 1983, A\&A, 118, 4
\bibitem{a2} Boulares A. \& Cox D.P. 1990, ApJ, 365, 544
\bibitem{a3} Cayatte V., van Gorkom J.H., Balkowski C., \& Kotanyi C. 1990, AJ, 100, 604
\bibitem{a4} Cayatte V., Kotanyi C., Balkowski C., \& van Gorkom J.H. 1994, AJ, 107, 1003 
\bibitem{a5} de Vaucouleurs G., de Vaucouleurs A., Corwin H.G., et al. 1991, Third Reference Catalogue of Bright Galaxies,(New York:Springer)(RC3)
\bibitem{a6} Elmegreen B.G. \& Falgaron E. 1996, ApJ, 471, 816
\bibitem{a7} Guharthakurta P., van Gorkom J.H., Kotanyi C.G., \& Balkowski C. 1988, AJ, 96, 851
\bibitem{a8} Keel W.C. 1996, PASP, 108, 917
\bibitem{a8a} Kenney J.D. \& Young J.S. 1988, ApJS, 66, 261
\bibitem{a9} Kenney J.D., van Gorkom J.H., \& Vollmer B. 2003, in preparation 
\bibitem{a10} Knapen J.H., Cepa J., Beckman J.E., Soledad del Rio M., \& Pedlar A. 1993, ApJ, 416, 563
\bibitem{a11} Napier P.J., Thompson A.R., \& Ekers R.D. 1983, Proc. IEEE 71, 1295
\bibitem{a12} Phookun B. \& Mundy L.G. 1995, ApJ, 453, 154
\bibitem{a13} Schulz S. \& Struck C. 2001, MNRAS, 328, 185
\bibitem{a14} Solanes J.M., Manrique A.,  Garcia-Gomez C., et al. 2001, ApJ, 548, 97
\bibitem{a15} Springel V., Yoshida N., \& White D.M. 2001, NA, 6, 79
\bibitem{a16} Tsch\"{o}ke D., Bomans D.J., Hensler G., \& Junkes N. 2001, A\&A, 380, 40
\bibitem{a17} van den Bergh S. 1976, ApJ, 206, 883
\bibitem{a18} Vollmer B., Cayatte V., Boselli A., Balkowski C., \& Duschl W.J. 1999, A\&A, 349, 411
\bibitem{a19} Vollmer B., Cayatte V., Balkowski C., \& Duschl W.J. 2001, ApJ, 561, 708
\bibitem{a20} Vollmer B. \& Huchtmeier W. 2003, A\&A, 406, 427
\bibitem{a21} Vollmer B. 2003, A\&A, 398, 525
\bibitem{a22} Warmels R.H. 1988, A\&AS, 72, 57 
\bibitem{a23} Wiegel W. 1994, Diploma Thesis, University of Heidelberg

\end{thebibliography}
\end{document}